\documentclass[aps,english,twocolumn,superscriptaddress]{revtex4-1}

\usepackage{longtable}
\usepackage{morefloats}
\usepackage[dvips]{graphicx}
\usepackage{color}
\usepackage{epsfig,graphicx,amsfonts,amsbsy}
\usepackage{amsmath,amsfonts,amsthm,amssymb}
\usepackage{appendix}
\usepackage{makeidx}
\usepackage{url}
\usepackage{verbatim}
\usepackage[bookmarksnumbered,pdfpagelabels=true,plainpages=false,colorlinks=true,linkcolor=blue,citecolor=red,urlcolor=blue]{hyperref}
\usepackage[rightcaption]{sidecap}
\usepackage{array}
\usepackage{booktabs}
\usepackage{multirow}
\usepackage{tabularx}
\usepackage{cancel,soul,ulem}

\newcommand{\bs}[1]{\boldsymbol{#1}}

\makeatletter

\begin{document}
\title{Many-body theory of spin-current driven instabilities in magnetic insulators}
\author{Roberto E. Troncoso}
\affiliation{Center for Quantum Spintronics, Department of Physics, Norwegian University of Science and Technology, NO-7491 Trondheim, Norway}
\email{r.troncoso@ntnu.no}
\affiliation{Institute for Theoretical Physics and Center for Extreme Matter and Emergent Phenomena,
Utrecht University, Leuvenlaan 4, 3584 CE Utrecht, The Netherlands}
\author{Arne Brataas}
\affiliation{Center for Quantum Spintronics, Department of Physics, Norwegian University of Science and Technology, NO-7491 Trondheim, Norway}
\author{Rembert A. Duine } 
\affiliation{Institute for Theoretical Physics and Center for Extreme Matter and Emergent Phenomena,
Utrecht University, Leuvenlaan 4, 3584 CE Utrecht, The Netherlands}
\affiliation{Department of Applied Physics, Eindhoven University of Technology, P.O. Box 513, 5600 MB Eindhoven, The Netherlands}

\begin{abstract}
We consider a magnetic insulator in contact with a normal metal. We derive a self-consistent Keldysh effective action for the magnon gas that contains the effects of magnon-magnon interactions and contact with the metal to lowest order. Self-consistent expressions for the dispersion relation, temperature and chemical potential for magnons are derived. Based on this effective action, we study instabilities of the magnon gas that arise due to spin-current flowing across the interface between the normal metal and the magnetic insulator. We find that the stability phase diagram  is modified  by an interference between magnon-magnon interactions and interfacial magnon-electron coupling. These effects persist at low temperatures and for thin magnetic insulators. 
\end{abstract}

%75.30.Ds---Magnetic properties, spin waves
%75.76.+j----Spin transport effects
%67.85.Hj----Bose-Einstein condensates

\maketitle

\section{Introduction}

Understanding the interplay between magnetization dynamics and spin-currents is a fundamental issue that is relevant for  spintronic devices \cite{Wolf,Takahashi,Arne1}. In particular, there is a growing interest, both theoretically \cite{Tserkovnyak,J.Xiao1,J.Xingtao,J.Xiao2,Y.T.Chen,Kapelrud,Y.Zhou,S.S.L.Zhang} and experimentally \cite{Y.Kajiwara,C.W.Sandweg,W.Sandweg,Z.Wang,S.M.Rezende,H.Nakayama,Z.Qiu}, in magnetic insulators such as Yttrium Iron Garnet (YIG) in contact with heavy metals like Pt (YIG$|$Pt bilayers). In these hybrid systems, magnons and magnetization dynamics are excited via  interfacial spin-transfer torques \cite{Berger}. The realization of a Bose-Einstein condensate (BEC) of magnons  through this mechanism has recently been proposed \cite{Bender}. Auto-oscillations driven by the spin Hall effect \cite{Klein2014} and thermal spin current \cite{Safranski2016,Lauer2016} have very recently been observed. Earlier, a condensate has been realized  at room temperature in YIG by other means, namely via parametric pumping \cite{Demokritov2006}. This is an example of a non-equilibrium condensate of quasiparticles \cite{Rezende, Bugrij,Troncoso,Fli,Troncoso2}. Such non-equilibrium BECs have attracted a great deal of attention and occur in different physical systems such as excitons \cite{Eisenstein,Butov, Fukuzawa}, phonons \cite{Misochko}, polaritons \cite{Kasprzak} and photons \cite{Klaers}. Specifically, condensation of magnons has stimulated efforts to control coherent transport of spin waves at room temperature \cite{Kruglyak}.

\begin{figure}[ht]
\includegraphics[width=3.3in]{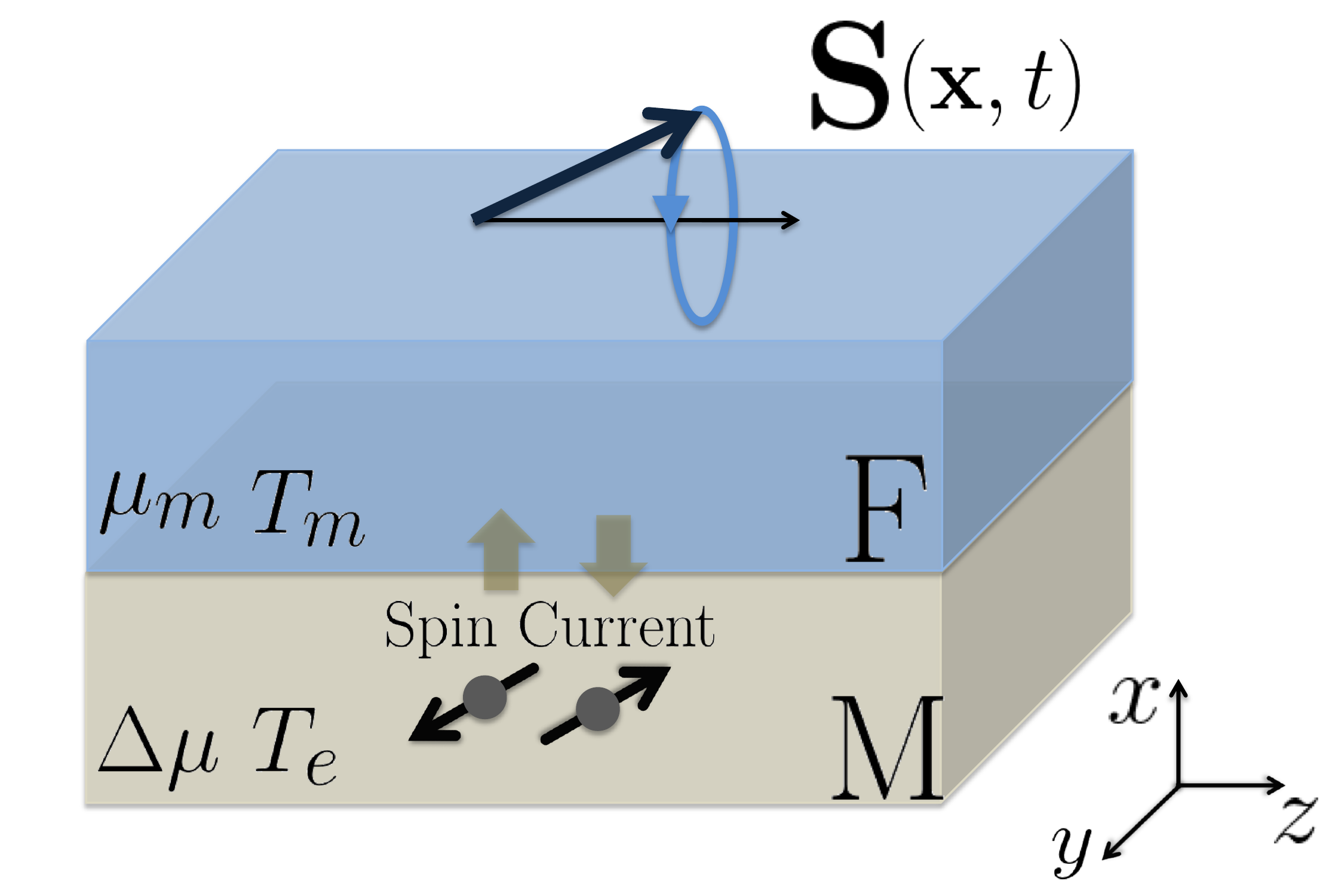}
\caption{Schematic illustration of spin-transfer induced spin-wave excitations of the insulating magnetic film in contact with a normal metal (F$|$M). A spin accumulation $\Delta \mu$ is induced in the metal via the spin Hall effect. The magnons are assumed to be in quasi-equilibrium with a chemical potential $\mu_m$. In turn, the spin accumulation exerts a torque on the magnetization in F. A temperature gradient $\Delta T = T_m-T_e$, applied longitudinally also contributes to the net spin-current flow through the interface.}
\label{fig:model}
\end{figure}

In this article, we present a microscopic study of magnon instabilities (such as Bose-Einstein condensation and/or swasing \cite{bender2014}) in insulating ferromagnets (F) induced via spin-current injection through the interface with an adjacent normal metal (M, see Fig. \ref{fig:model}). The interfacial spin-current is generated by the combined effects of a thermal gradient across the interface \cite{J.Xiao1,spincurrent} and the spin Hall effect in the normal metal (leading to a spin accumulation in the normal metal at the interface).  We derive a Keldysh effective action for the magnons in the ferromagnet up to second-order in the coupling with the normal metal. Through this approach, self-consistent relations are derived for the thermodynamic variables and the dispersion relation of magnons. We use this description to find the stability phase diagram. Besides introducing a new theoretical framework, this approach improves the previous treatment \cite{bender2014} by including interference effects between magnon-magnon interactions and interfacial magnon-electron coupling. These effects are finite at low temperature, and may prevent instabilities if they are very strong.

The remainder of this article is organized as follows. The following Sec. \ref{sec2} introduces our model to describe the magnon dynamics and its coupling to electrons. In Sec. \ref{sec3}, we proceed to derive the effective action for the magnons within the functional formulation of the Schwinger-Keldysh formalism. In Sec. \ref{sec4}, we construct the stability phase diagram. We end in Sec. \ref{sec5} by summarizing our results with a brief discussion and conclusion. In the appendices, we detail various technical steps of the calculations.

%%%%%
\section{Model}\label{sec2}
%%%%%
The system is a ferromagnet in contact with a normal metal as in Fig. \ref{fig:model}. We  consider a three-dimensional system of localized spins in quasi-equilibrium at temperature $T_m$ and magnon chemical potential $\mu_m$ \cite{Cornelissen2016,Du2016}. The normal metal is at temperature $T_e$ and has a spin accumulation $\Delta \mu$.

The insulating ferromagnet consists of spins ${\bf S}_{{\bf x}}$ located at position ${\bf x}$,
%=\left\{j,{\bf r}\right\}$. The $j-$coordinate is along the $x-$axis and ${\bf r}$ 
being the F$|$M interface parallel to the $xy-$plane (see Fig. \ref{fig:model}). The magnetic system is described by the spin Hamiltonian 
\begin{align}
{\cal H}_{\text{S}}=-J_{xc}\sum_{\langle {\bf x},{\bf x}'\rangle }\hat {\bf S}_{{\bf x}}\cdot\hat {\bf S}_{{\bf x}'}-\sum_{{\bf x}}{\bf B}\cdot\hat {\bf S}_{{\bf x}}+\frac{K_{z}}{2}\sum_{\bf x}(\hat {S}^z_{\bf x})^2,
\end{align}
where $J_{xc}$ is the exchange coupling between nearest neighbors as indicated by the notation $\langle \cdot, \cdot \rangle$, $K_z$ is an easy-plane anisotropy constant and ${\bf B}=B \hat z$ is the external magnetic field in units of energy that points in the $z$-direction. We consider linearized spin excitations, magnons, around the $z$-direction for sufficiently large fields. These are introduced by the Holstein-Primakoff  transformation \cite{Holstein,Akhiezer,Auerbach} that quantizes the spins in terms of bosons, $\hat {S}^+_{\bf x}=\sqrt{2S-{\hat{b}^{\dagger}_{\bf x}\hat{b}_{\bf x}}} \hat{b}_{\bf x}$, $\hat {S}^-_{\bf x}=\hat{b}^{\dagger}_{\bf x}\sqrt{2S-{\hat{b}^{\dagger}_{\bf x}\hat{b}_{\bf x}}}$ and $\hat { S}_z=\left(S-{\hat b}^{\dagger}_{{\bf x}}{\hat b}_{{\bf x}}\right)$  with $S$ the spin quantum number. One magnon, created (annihilated) at site ${\bf x}$ of the lattice by the operator ${\hat b}^{\dagger}_{{\bf x}}$(${\hat b}_{{\bf x}}$), corresponds to changing the total spin with $+\hbar$($-\hbar$). Expanding up to fourth order in the magnon operators, the spin Hamiltonian becomes ${\cal H}_{\text{S}} [\hat b, \hat b^\dagger]\simeq{\cal H}^{(2)}_{\text{S}} [\hat b, \hat b^\dagger]+{\cal H}^{(4)}_{\text{S}} [\hat b, \hat b^\dagger]$, where terms up to order $1/\sqrt{S}$ are kept. In momentum space the first part of the Hamiltonian is given by 
\begin{align}\label{eq:freemagnons}
{\cal H}^{(2)}_{\text{S}} [\hat b, \hat b^\dagger]=\sum_{{\bf q}}\epsilon_{\bf q}{\hat b}^{\dagger}_{{\bf q}}{\hat b}_{{\bf q}}.
\end{align}
In the long wavelength limit, the magnon dispersion is $\epsilon _{\bf q}=Aq^2+\epsilon_0$, where $\epsilon_0=B-SK_z$ is the magnon gap, $A=3SJ_{xc}a^2$ is the spin stiffness and $a$ is the lattice constant. Furthermore, 
\begin{align}\label{eq:mminteraction}
{\cal H}^{(4)}_{\text{S}} [\hat b, \hat b^\dagger]=\sum_{{\bf q}_1,{\bf q}_2,{\bf q}_3,{\bf q}_4}V^{(2,2)}_{{\bf q}_1,{\bf q}_2,{\bf q}_3,{\bf q}_4}{\hat b}^{\dagger}_{{\bf q}_1}{\hat b}^{\dagger}_{{\bf q}_2}{\hat b}_{{\bf q}_3}{\hat b}_{{\bf q}_4}.
\end{align}
This part of the Hamiltonian represents magnon-magnon interactions that result from exchange and anisotropy that causes the thermalization of magnons in the ferromagnet \cite{Akhiezer}. We approximate the scattering amplitude by $V^{(2,2)}_{{\bf q}_1,{\bf q}_2,{\bf q}_3,{\bf q}_4}=\frac{K_z}{2N}\delta_{{\bf q}_1+{\bf q}_2-{\bf q}_3-{\bf q}_4}$, and thus neglect the contributions from the exchange interactions. This approximation is valid for magnons with sufficiently long wavelengths and is for our purposes justified as we are interested in long wavelength instabilities in the ferromagnet. 

The electronic degrees of freedom in the metal are described by the tight-binding Hamiltonian
\begin{align}\label{eq:electrons}
{\cal H}_{e}=-t \sum_{\langle {\bf x},{\bf x}'\rangle,\sigma}{\hat\psi}^{\dagger}_{{\bf x},\sigma}{\hat\psi}_{{\bf x}',\sigma}-\sum_{{\bf x},\sigma}\mu_{\sigma}{\hat\psi}^{\dagger}_{{\bf x},\sigma}{\hat\psi}_{{\bf x},\sigma},
\end{align}
in terms of second-quantized operators ${\hat\psi}^{\dagger}_{{\bf x},\sigma}$ and its hermitian conjugate that create, respectively, annihilate an electron, and with $t$ the hopping amplitude. We also include a spin-dependent chemical potential $\mu_{\sigma}$. The latter results from the spin Hall effect in the normal and defines a nonzero spin accumulation $\Delta \mu= \mu_\uparrow-\mu_\downarrow$. 

We assume the magnons and electrons predominantly interact via an exchange coupling between the spin density of electrons and localized magnetic moments facing the F$|$M interface. The Hamiltonian that couples metal and insulator is then
\begin{align}\label{eq:electronmagnoncoupling}
{\cal H}_{e-m}=-\sum_{{\bf x},{\bf x}'}J_{{\bf x}{\bf x}'}\hat {\bf{S}}_{{\bf x}}\cdot{\bf{\hat s}}_{{\bf x}'},
\end{align}
where the coupling strength $J_{{\bf x}{\bf x}'}=J_{jj'}\delta_{{\bf r},{\bf r}'}$, depends on the details of the interface and is assumed to be local. The spin-density of electrons at  site ${\bf x}$ is ${\bf{\hat s}}_{\bf x}=\sum_{\sigma,\sigma'}\hat{\psi}^{\dagger}_{{\bf x},\sigma}\bs{\tau}_{\sigma\sigma'}\hat{\psi}_{{\bf x},\sigma'}$, with $\bs{\tau}$ the Pauli matrix vector. After the Holstein-Primakoff transformation on the spins in the insulator, the  electron-magnon Hamiltonian is to quadratic order in the magnon operators
\begin{align}\label{eq:electronmagnoncouplingsq}
{\cal H}_{e-m}\nonumber=&-\sum_{{\bf x},{\bf x}'}J_{{\bf x}{\bf x}'}\left[\sqrt{2S}\left({\hat b}^{\dagger}_{{\bf x}}{\hat\psi}^{\dagger}_{{\bf x}',\downarrow}{\hat\psi}_{{\bf x}',\uparrow}+h.c.\right)\right.\\
&\left.+\left(S-{\hat b}^{\dagger}_{{\bf x}}{\hat b}_{{\bf x}}\right)\left({\hat\psi}^{\dagger}_{{\bf x}',\uparrow}{\hat\psi}_{{\bf x}',\uparrow}-{\hat\psi}^{\dagger}_{{\bf x}',\downarrow}{\hat\psi}_{{\bf x}',\downarrow}\right)\right].
\end{align}

Hence, when an electron flips its spin at the interface it creates (or annihilates) one magnon in the insulating ferromagnet \cite{Tserkovnyak}.

%%%%%%%%%%%%%%%%%%%%%%%%%%%%%%%%%%%%%%%%%%%
\section{Nonequilibrium theory}\label{sec3}
%%%%%%%%%%%%%%%%%%%%%%%%%%%%%%%%%%%%%%%%%%%
In this section, we derive an effective action for the magnon gas using a functional Keldysh approach \cite{stoof1}. We also calculate the self-energy that magnons acquire by their interaction with the electrons.
%%%%%%%%%%%%%%%%%%%%%%%%%%%%%%%%%%%%%%%%%%%%%%%%%%%%%%%%
\subsection{Self-energy due to electron-magnon coupling}
\begin{figure}[ht]
\begin{center}
\includegraphics[width=3.8in]{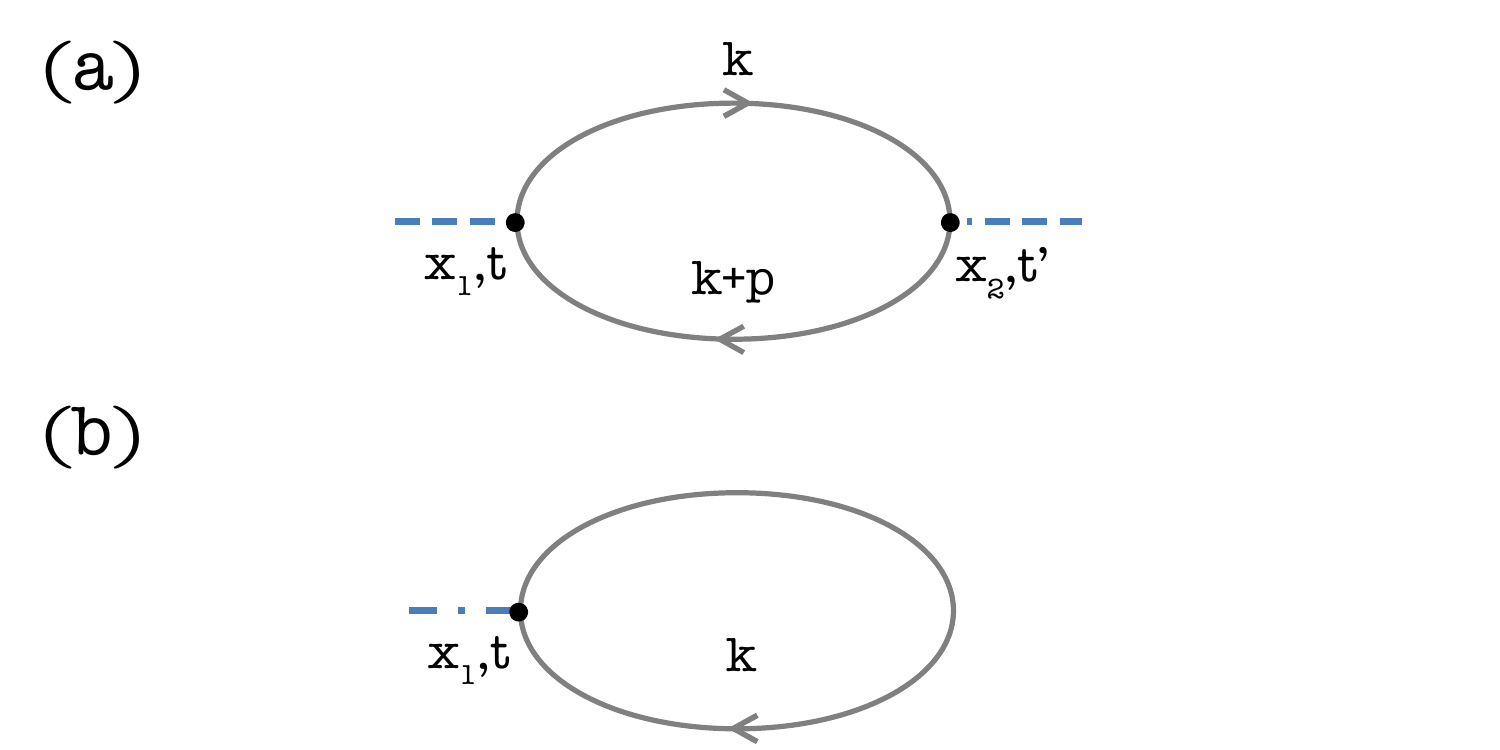}
\caption{Feynman diagrams contributing to the magnon dynamics due magnon-electron interactions at the F$|$M interface. (a) Electron bubble diagram representing the annihilation and creation of magnons via spin-flip processes. (b) Diagram for the enhanced spin polarization. The momentum carried by the magnon is denoted by $\bf p$.}
\label{fig:electronBubble}
\end{center}
\end{figure}

%%%%%%%%%%%%%%%%%%%%%%%%%%%%%%%%%%%%%%%%%%%%%%%%%%%%%%%%
The starting point is the functional integral 
\begin{align}\label{eq:partitionfunction}
{\cal Z}=\int {\cal D}\phi^*{\cal D}\phi{\cal D}\psi^*{\cal D}\psi  \text{exp}\left\{\frac{i}{\hbar}{\cal S}[\psi,\psi^*,\phi,\phi^*]\right\}.
\end{align}
The action is expressed in terms of bosonic fields $\phi_{{\bf x}}(t)$, describing magnons, and fermionic fields $\psi_{{\bf x},\sigma}(t)$, describing electrons
\begin{align}\label{eq:totalaction}
\nonumber{\cal S}[\psi,&\psi^{*},\phi,\phi^{*}]=\int_{{\cal C}^{\infty}}dt\left\{\sum_{{\bf x}}\phi^{*}_{{\bf x}}(t)i\hbar\frac{\partial}{\partial t}\phi_{{\bf x}}(t)-{\cal{ H}}_{S}\left[\phi,\phi^{*}\right]\right.\\
&\left.+\int_{{\cal C}^{\infty}}dt'\sum_{{\bf x}{\bf x}'}\sum_{\sigma\sigma'}{\psi}^{*}_{{\bf x},\sigma}(t){\cal K}^{{\bf x}{\bf x}'}_{\sigma\sigma'}(t,t'){\psi}_{{\bf x}',\sigma'}(t')\right\},
\end{align}
where the magnon Hamiltonian is in the continuum limit given by
\begin{align}{\cal{H}}_S[\phi,\phi^*]=\int d{\bf x}\phi^*\left(-A\nabla^2+B-SK_z+\frac{K_z}{2}\left|\phi\right|^2\right)\phi,
\end{align} 
and follows directly from evaluating the Hamiltonian in Eqs. (\ref{eq:freemagnons}) and (\ref{eq:mminteraction}) for the bosonic fields and taking the continuum limit. 
The time integration in Eq. (\ref{eq:totalaction}) is  over the Keldysh contour ${\cal C}^{\infty}$, whereby the functional integral in Eq. (\ref{eq:partitionfunction}) is over all fields $\phi_{{\bf x}}(t)$ and $\psi_{{\bf x},\sigma}(t)$ that evolve forward in time from $-\infty$ to $t_0$, and backwards from $t_0$ to $-\infty$. The kernel in Eq. (\ref{eq:totalaction}) is defined as
\begin{align}\label{eq:kernel}
{\cal K}^{{\bf x}{\bf x}'}_{\sigma\sigma'}(t,t')\equiv{ G}^{-1}_{{\bf x}{\bf x}';\sigma\sigma'}(t,t')\delta(t,t')+\delta{\cal K}^{{\bf x}{\bf x}'}_{\sigma\sigma'}(t,t'),
\end{align}
where ${G}$ is the free Green's function for the electrons, that obeys 
\begin{align*}
\sum_{{\bf x}''}\left[i\hbar\frac{\partial}{\partial t}\delta_{{\bf x},{\bf x}''}+\text{t}_{{\bf x},{\bf x}'';\sigma}\right]{G}_{{\bf x}''{\bf x}';\sigma\sigma'}(t,t')=\hbar\delta(t,t')\delta_{\sigma,\sigma'}\delta_{{\bf x},{\bf x}'},
\end{align*}
with $\text{t}_{{\bf x},{\bf x}';\sigma}=\left[\text{t}\left(\delta_{{\bf x},{\bf x}'+{\bf 1}}+\delta_{{\bf x},{\bf x}'-{\bf 1}}\right)+\mu_{\sigma}\delta_{{\bf x},{\bf x}'}\right]$, where the notations $\pm {\bf 1}$ denotes all nearest neighbours. The interactions between electronic spin and localized magnetic moments are described by 
\begin{align}\label{eq:deltakernel}
\delta{\cal K}^{{\bf x}{\bf x}'}_{\sigma\sigma'}(&t,t')\nonumber=\sqrt{2S}\delta_{{\bf x}{\bf x}'}\sum_{{\bf x}''}J_{{\bf x}''{\bf x}'}\left[\phi^{*}_{{\bf x}''}(t)\tau^{+}_{\sigma\sigma'}+\phi_{{\bf x}''}(t)\tau^{-}_{\sigma\sigma'}\right.\\
&\left.
+\frac{1}{\sqrt{2S}}\left(S-\phi^{*}_{{\bf x}''}(t)\phi_{{\bf x}''}(t)\right)\tau^{z}_{\sigma\sigma'}\right]\delta(t,t'),
\end{align}
where $\tau^{\pm}=\left(\tau^x\pm i\tau^y\right)/2$. We now integrate out the electronic degrees of freedom $\psi_{{\bf x},\sigma}(t)$ in Eq. (\ref{eq:partitionfunction}). The functional integration can be done exactly since the action is a Gaussian integral, see Eq. (\ref{eq:totalaction}), which lead us to an effective theory for the magnons in the ferromagnet,
\begin{align}\label{eq:partitionfunction1}
{\cal Z}=\int {\cal D}\phi^*{\cal D}\phi \text{exp}\left\{\frac{i}{\hbar}{\cal S}_{\text{m}}[\phi^*,\phi]\right\}~.
\end{align}
where ${\cal S}_{\text{m}}$  contains all the electronic information and its influence on the magnon dynamics. To second order in the electron-magnon coupling, we obtain for the effective action 
\begin{align}
{\cal S}_{\text{m}}\left[\phi,\phi^{*}\right]=&\int_{{\cal C}^{\infty}}dt\left\{\sum_{{\bf x}}\phi^{*}_{{\bf x}}(t)i\hbar\frac{\partial}{\partial t}\phi_{{\bf x}}(t)-{\cal{ H}}_{S}\left[\phi,\phi^{*}\right]\right\} \nonumber \\
&-i{\text{Tr}}\left[{ G}\delta{\cal K}\right]+i{\text {Tr}}\left[{ G}\delta{\cal K}{ G}\delta{\cal K}\right]/2\hbar~,
\end{align}
where the two last terms correspond to electron-magnon scattering processes described by the Feynman diagrams in Fig. \ref{fig:electronBubble}. These processes dresses the magnons with the self-energy  
\begin{align}\label{eq:selfenergygeneralized}
\hbar\Sigma_{\text{st};{\bf x}{\bf x}'}(t,t')=\nonumber 2&S\sum_{{\bf x}_1{\bf x}_2}J_{{\bf x}{\bf x}_1}J_{{\bf x}'{\bf x}_2}\Pi^{\uparrow\downarrow}_{{\bf x}_1{\bf x}_2}(t,t')\\
&-\delta_{{\bf x},{\bf x}'}\delta(t,t')\sum_{{\bf x}_1}J_{{\bf x}_1{\bf x}'}\delta{\bar n}({\bf x}_1,t),
\end{align}
where the electron bubble reads $\Pi^{\sigma\sigma'}_{{\bf x}{\bf x}'}(t,t')=-\frac{i}{\hbar} G_{{\bf x}{\bf x}';\sigma}(t,t')G_{{\bf x}'{\bf x};\sigma'}(t',t)$, which contains the information of the electronic quasi-particles in the metal. The first term on the right-hand side accounts for the creation and annihilation of magnons by means of a spin-flip process, see Fig. \ref{fig:electronBubble} (a). The second term corresponds to the effective field produced by the imbalance of the electronic density at the interface, $\delta{\bar n}={\bar n}_{\uparrow}-{\bar n}_{\downarrow}$, where ${\bar n}_{\sigma}=n_{\sigma}-S\sum_{{\bf x}_2,{\bf x}_3}\int_{\cal C}dt'J_{{\bf x}_3{\bf x}_2}\Pi^{\sigma\sigma}_{{\bf x}_1{\bf x}_2}(t,t')\tau^{z}_{\sigma\sigma}$, see Fig. \ref{fig:electronBubble} (b). 
%%%%%%%%
%%%%%%%%Picture of bubble diagrams
%%%%%%%%

Before we proceed to derive an effective action for the long-wavelength dynamics, we first make some simplifying assumptions. First, we assume that the ferromagnetic film is very thin so that spin-waves only propagate parallel to the interface. After a Fourier transform the action simplifies to 
\begin{align}\label{eq:effectiveaction}
{\cal S}_{\rm em}\left[\phi,\phi^{*}\right]\nonumber=&\int_{{\cal C}^\infty}dt \int \frac{d{\bf q}}{(2\pi)^2}\left(\phi^{*}_{{\bf q}}(t)i\hbar\frac{\partial}{\partial t}\phi_{{\bf q}}(t)-{\cal H}_{\text{S}}\left[\phi,\phi^{*}\right]\right.\\
&\left.-\int_{{\cal C}^\infty}dt'\phi^{*}_{{\bf q}}(t)\hbar\Sigma_{\text{st};{\bf q}}(t,t')\phi_{{\bf q}}(t')\right)~,
\end{align} 
where ${\bf q}$ is the in-plane wavevector, i.e., along  the interface (see Fig. \ref{fig:model}). The self-energy is written in momentum space, where $\hbar\Sigma_{\text{st};{\bf q}}=\sum_{k_{\perp}}\hbar\Sigma_{\text{st};{\bf k}}$, with the latter summation over all electronic wavevectors perpendicular to the interface. 

%%%%%%%%Picture of the bubble Tmatrix
\begin{center}
	\begin{figure}[h!]
		\includegraphics[width=3.4in]{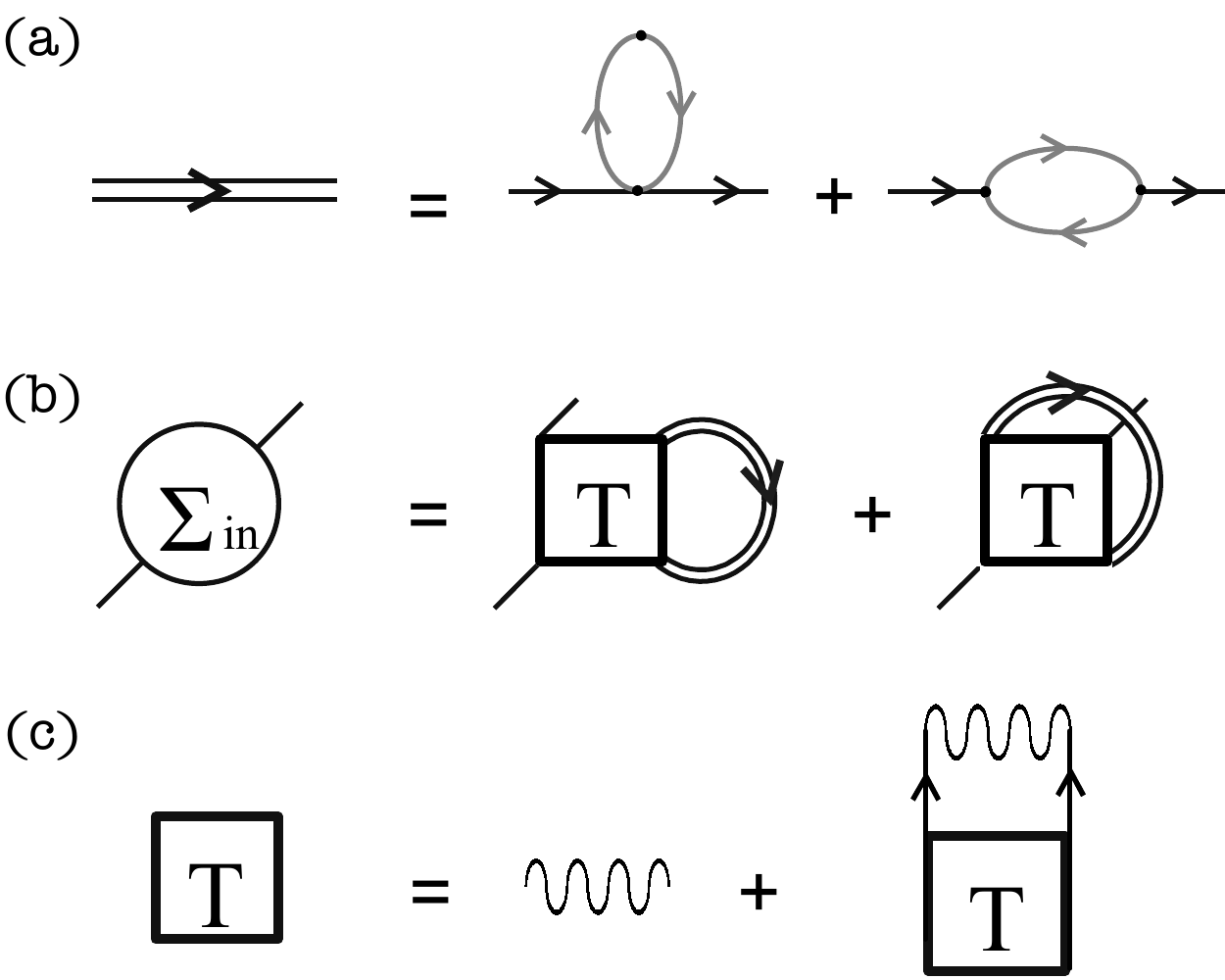}
		\caption{(a) Feynman diagram for the dressed Green's function of magnons by the self-energy Eq. (\ref{eq:selfenergygeneralized}), due to the coupling with the normal metal. (b) Diagrammatic representation for the interacting self-energy and (c) the T-matrix ladder approximation, where the wiggly lines denotes the contact interaction $K_z$. Note that the self-energy and T-matrix are determined by the dressed and free propagators, respectively.%\textcolor{red}{grey part should be adapted in same way as previous Feynman diagram figure}
		}
		\label{fig:DiagrammaticRepresentation}
	\end{figure} 
\end{center}
%%%%%%%%
Second, we assume that the coupling occurs at the interface only, so that $J_{{\bf x}{\bf x}'}=J\delta_{jj'}\delta_{{\bf r},{\bf r}'}$. Then the retarded component of this self-energy, which we need later on, is (see Appendix \ref{sec:SelfEnergy}) 
\begin{align}
\hbar\Sigma^{(\pm)}_{\text{st};{\bf x}{\bf x}'}(t,t')= 2&S\sum_{{\bf x}_1{\bf x}_2}J_{{\bf x}{\bf x}_1}J_{{\bf x}'{\bf x}_2}\Pi^{\uparrow\downarrow,(\pm)}_{{\bf x}_1{\bf x}_2}(t,t'),
\end{align}
where the Fourier transform of the electron bubble, in energy and momentum space, is
\begin{align}\label{eq:selfenergyRA}
\Pi^{\uparrow\downarrow}{}^{,(\pm)}({\bf k},\epsilon)\nonumber=&\frac{1}{2}N(0)a^3\left[\frac{1}{3}\left(\frac{{\bf k}}{2k_F}\right)^2-1\right]-\frac{N(0)a^3\epsilon\Delta\mu}{16\epsilon^2_F}\\
&\nonumber\times\left[\left(\frac{2k_F}{\bf k}\right)^2+1\right]\pm i \frac{\pi N(0)a^3k_F}{8\epsilon_F\left|{\bf k}\right|}\left(\Delta\mu-\epsilon\right)\\
&\qquad\qquad\qquad\times\Theta\left[1-\left(\frac{|{\bf k}|}{2k_F}\right)^2\right],
\end{align}
with $N(0)$ the electronic density of states at the Fermi level, and $k_F$ and $\epsilon_F$ the Fermi wave number and energy, respectively.

%%%%%%%%%%%%%%%%%%%%%%%%%%%%%
\subsection{Effective action}
%%%%%%%%%%%%%%%%%%%%%%%%%%%%%
We are interested in real-time dynamics of the magnon gas, thus we expect the effective action to depend only on the retarded part of $\hbar\Sigma_{\text{st};{\bf q}}$. Moreover, since the magnon-magnon interactions resulting from the anisotropy are short-ranged, we can use the so-called ladder approximation. We project, in Eq. (\ref{eq:effectiveaction}), the fields onto the real-time axis by the substitution $\phi_{\pm}= \psi \pm \xi/2$, with $\psi$ the classical field and where $\pm$ refer to the upper and lower branch of the Keldysh contour. The field $\xi$ denote the quantum fluctuations which by the current purpose are disregarded. Details of this calculation are outlined in Appendix \ref{sec:LadderApproximation}. Following Ref. \cite{stoof1}, we ultimately find a one-particle-irreducible long-wavelength effective action for the classical field $\psi$ that is given by
\begin{align}\label{eq:effAction}
{\cal S}_{\text{eff}}\left[\psi,\psi^{*}\right]\nonumber&=\int dt\int \frac{d{\bf q}}{(2\pi)^2}\psi^{*}({\bf q},t)\left(i\hbar\frac{\partial}{\partial t}-\epsilon'\left({\bf q}\right)+\mu_m\right.\\
&\left.-{a^2}K_zn_{\text{th}}-\frac{K_z}{2}\left|\psi({\bf q},t)\right|^2\right)\psi({\bf q},t).
\end{align}

The single-particle dispersion relation of magnons is renormalized by the magnon-magnon interactions and the coupling with electrons, and obeying the self-consistent relation 
\begin{align}\label{eq:singleenergyrenormalized}
\epsilon'\left({\bf q}\right)=\epsilon\left({\bf q}\right)+\text{Re}\left[\hbar\Sigma^{(+)}({\bf q};\epsilon'({\bf q})-\mu_m)\right].
\end{align}

Here, $\hbar\Sigma=\hbar\Sigma_{\text{st}}+\hbar\Sigma_{\text{in}}$, is the sum of contributions due to the coupling with electrons and magnon-magnon interactions, and $\mu_m$ the chemical potential of magnons. Moreover, we have included in the action Eq. (\ref{eq:effAction}) a mean-field interaction with the thermal cloud of magnons, whose density is  $n_{\text{th}}=\zeta\left(3/2\right)d_F\left(k_BT_m/4\pi A\right)^{3/2}$, with $d_{\text{F}}$ the thickness of ferromagnet. The self-energy $\hbar\Sigma_{\text{in}}$ due to interactions is diagrammatically shown in Fig. \ref{fig:DiagrammaticRepresentation}. The real part of $\hbar\Sigma_{\text{in}}$ --- which we need later on --- reads 
\begin{widetext} 
\begin{align}\label{eq:IntselfenergyFinal}
\text{Re}\left[\hbar\Sigma^{(+)}_{\text{in}}({\bf 0},\epsilon_0-\mu_m))\right]=-{a^2}{K_z}\int\frac{d{\bf q}'}{(2\pi)^2}&\int\frac{d\epsilon'}{(2\pi)}\nonumber N_B(\beta_e(\epsilon'-\Delta\mu))\frac{\text{Im}\left[\hbar\Sigma^{(+)}_{\text{st}}({\bf q}',\epsilon')\right]}{\left(\epsilon'-\epsilon({\bf q}')+\mu_m\right)^2}\\
	-\frac{{a^2}K^2_z}{4A}&\int\frac{d{\bf q}'}{(2\pi)^2}\int\frac{d{\bf q}''}{(2\pi)^2}N_B(\beta_m (\epsilon({\bf q}')-\mu_m))N_B(\beta_m (\epsilon ({\bf q}'')-\mu_m))\frac{{\cal P}}{{\bf q}'\cdot{\bf q}''},
	\end{align}
\end{widetext}
with the first term on the right-hand side representing the interference between magnons and electrons. This can be seen in Fig. \ref{fig:DiagrammaticRepresentation}(a) and (b) where magnons, and thus their interactionss, are dressed by their coupling with electrons. Second term corresponds to the usual shift of the energy due to the two-particle interaction. To obtain Eq. (\ref{eq:IntselfenergyFinal}) it has been assumed that the momentum $\hbar{\bf q}$ is much smaller than the thermal momenta $\hbar/\Lambda_{\text{th}}$ \cite{stoof1}, with $\Lambda_{\text{th}} = \sqrt{4 \pi A/k_B T_m}$ the thermal de Broglie wavelength for the magnons. In the above, $N_B (x) = [e^x-1]^{-1}$ is the Bose distribution function, $\beta_e=1/k_B T_e$ is the inverse thermal energy of the electrons, and $\beta_m=1/k_B T_m$ that of the magnons.

When the energy of the single-particle state $\epsilon'\left({\bf q}\rightarrow {\bf 0}\right)$ becomes less than $\mu_m-{a^2}K_zn_{\text{th}}$, the magnon system becomes unstable. This signals the formation of a magnon Bose-Einstein condensate, a precessional instability, or magnetization reversal. The criterion for such an instability is thus
\begin{align}\label{eq:magnonBECcriterion}
\epsilon'\left({\bf q}\rightarrow {\bf 0}\right)+{a^2}K_zn_{\text{th}}-\mu_m<0.
\end{align}
Based on this condition, a phase diagram is determined in the next section. It is worth to comment that Eq. (\ref{eq:magnonBECcriterion}) involve self-consistent physical quantities such as the magnon energy, Eq. (\ref{eq:singleenergyrenormalized}), magnon temperature and chemical potential. Unlike previous works \cite{Bender,bender2014}, all these quantities can be evaluated self-consistently at leading order in the interfacial coupling $J$ with the electrons as we outline below.
%%%%%%%%Picture of magnon Chemical pot. and Temperature
\begin{figure}[h]
	\includegraphics[width=3.27in]{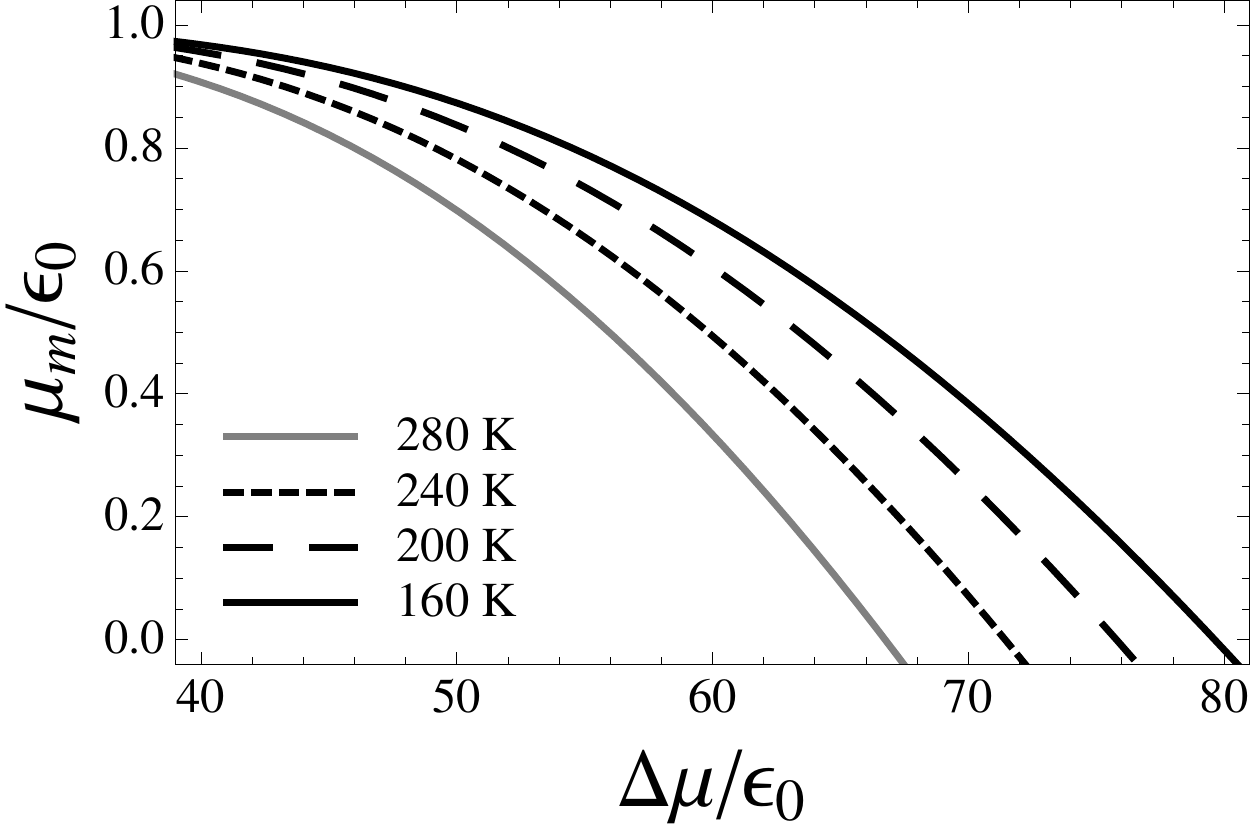}\vspace{-1.05cm}\\
	\includegraphics[width=3.29in]{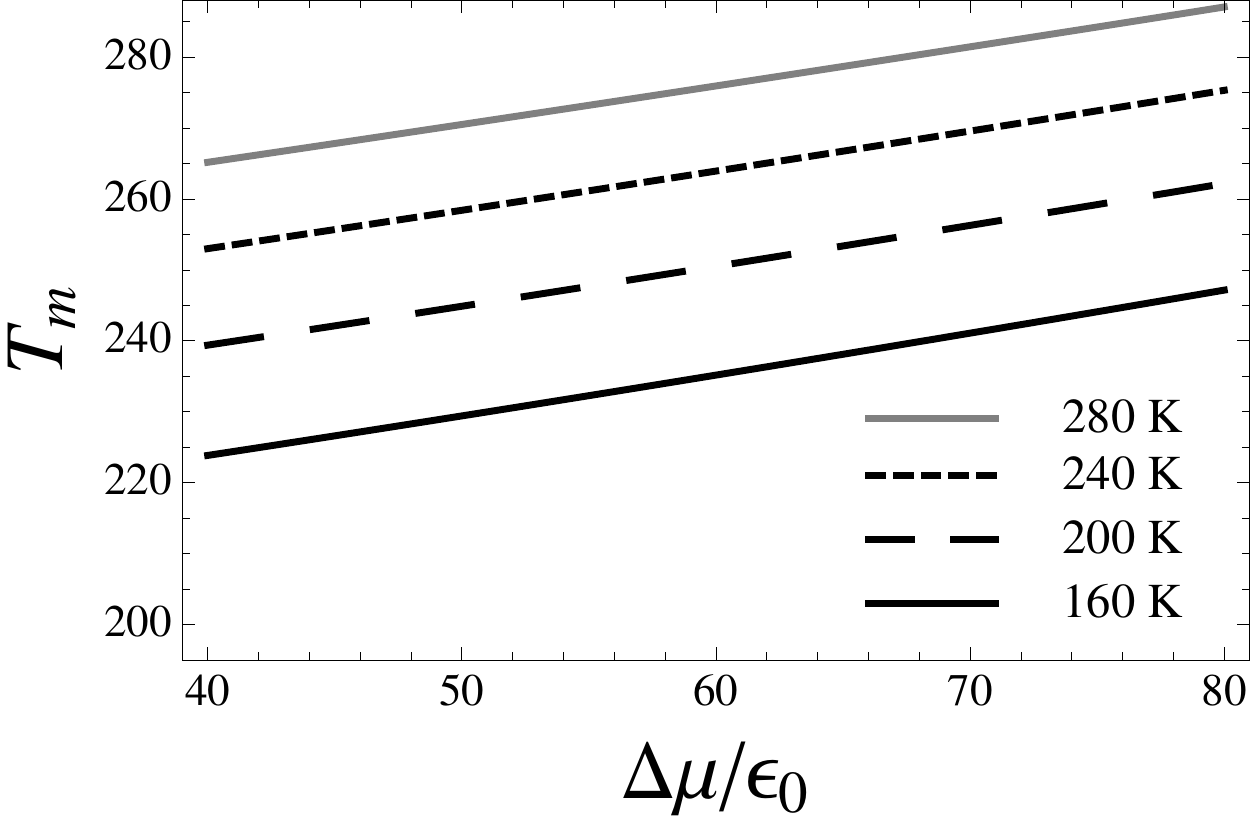}
	\caption{Chemical potential (top panel) and temperature of magnons (bottom panel) in the steady-state as a function of spin accumulation. The plots are displayed for different values of the temperature of the electrons.}
	\label{fig:selfconsistencyfigure}
\end{figure}
%%%%%%%%

Before proceeding to evaluate Eq. (\ref{eq:magnonBECcriterion}), however, we need to determine the magnon chemical potential and temperature for a given electron spin accumulation and temperature. This is done through the Boltzmann equation for magnons with the metallic coupling acting as a electronic reservoir that transfers spin and energy.  Details of this calculation are outlined in Appendix \ref{sec:SelfConsistent}. A Gilbert damping constant $\alpha$, parameterizing the coupling with phonons, is phenomenologically added. Finally, a steady state is required in the kinetic equations for the total density of magnons and total energy. In this limit, we find relations at thermodynamic equilibrium for the magnon temperature $T_m$ and chemical potential $\mu_m$ in terms of the temperature $T_e$ and spin accumulation $\Delta \mu$ of the electrons. These read
\begin{align}
(1+\chi)\text{Li}_{2}&\label{eq:criticalSAandT(a)}(z)+\frac{\epsilon_0(\chi-\bar\mu)-\Delta\mu}{k_B T_m}\text{Li}_{1}(z)=\frac{\bar\zeta\epsilon_0 T_e}{k_B T^2_m},\\
\nonumber\frac{1+\chi}{2}\text{Li}_{3}&(z)+\frac{{\epsilon_0(2\chi-\bar\mu)-\Delta\mu}}{k_B T_m}\text{Li}_{2}(z)\\
&\label{eq:criticalSAandT(b)}+\frac{\epsilon_0\left[\epsilon_0(\chi+\bar\mu)-\Delta\mu\right]}{k^2_B T^2_m}\text{Li}_{1}(z)=\frac{\bar\zeta^2\epsilon^2_0 T_e}{6k^2_BT^3_m},
\end{align}
where we used the PolyLogarithm function $\text{Li}_s(z)=\frac{1}{\Gamma(s)}\int^{\infty}_{0}dy\frac{y^{s-1}}{e^{y}z^{-1}-1}$, with $\Gamma(s)$ the gamma function and $z=e^{(\mu_m-\epsilon_0)/k_BT_m}$. Eqs. (\ref{eq:criticalSAandT(a)}) and (\ref{eq:criticalSAandT(b)}),
are solved for the quantities $\mu_m=\mu_m\left(\Delta\mu,T_e\right)$ and $T_m=T_m\left(\Delta\mu,T_e\right)$, with the following dimensionless parameters $\bar\mu=(\mu_m-\epsilon_0)/\epsilon_0$, $\chi=\frac{8\alpha\epsilon_F}{ k_FSJ^2N(0)a^4}$ and $\bar\zeta={8 A k_F^2}/{\epsilon_0}$, under consideration.

{The steady-state result for $\mu_m$ and $T_m$ is shown in Fig. \ref{fig:selfconsistencyfigure} as a function of spin accumulation and various temperatures $T_e$. These plots were determined for typical values of physical parameters, namely $\epsilon_0\approx 50 $~$\mu$eV, $S=1$ and $\alpha=10^{-5}$. Through the standard value for the spin-mixing conductance in the YIG$|$Pt interface, $g^{\uparrow\downarrow}=10^{14}\text{cm}^{-2}$ \cite{Y.Kajiwara,Heinrich}, we estimate the interfacial exchange coupling to be $J\approx50\text{meV}$ \cite{Heinrich}.}

%%%%%%%%%%%%%%%%%%%%%%%%%%%%%%%%%%%%%%%%%%%%%%%%%%%%%%%
\section{Condensed and normal phases of the magnon gas}\label{sec4}
%%%%%%%%%%%%%%%%%%%%%%%%%%%%%%%%%%%%%%%%%%%%%%%%%%%%%%%
In order to obtain results, we first rewrite the instability criterion in Eq. (\ref{eq:magnonBECcriterion}) to 
\begin{align}\label{eq:result_phasediagram}
\frac{\epsilon'\left({\bf q}\rightarrow {\bf 0}\right)+{a^2}K_zn_{\text{th}}-\mu_m}{(JN(0)a^3)k_BT_e}\equiv F[\delta\mu,\eta;e_0,e_1,e_2]<0,
\end{align}
where the effective chemical potential and the ratio between magnons and electrons temperature, have been defined as $\delta\mu=\left(\Delta\mu+\mu_m-\epsilon_0\right)/k_BT_e$ and $\eta=T_m/T_e$, respectively. {It should be noted that to get Eq. (\ref{eq:result_phasediagram}), we have used the strictly increasing and positive property of the first term on the right-hand side of Eq. (\ref{eq:IntselfenergyFinal}) as a function of $\Delta\mu$}. Moreover, the dimensionless function $F$ introduced above is given by $F[\delta\mu,\eta;e_0,e_1,e_2]=e_0+e_1\eta^{3/2}-e_2\delta\mu N_B(-\delta\mu)-\delta\mu$, with the dimensionless parameters obeying
\begin{align}
e_0&\nonumber=\frac{\epsilon_0-\mu_m}{k_BT_e}\left(\frac{1}{JN(0)a^3}-1\right)+\frac{2JS}{k_BT_e}\\
-&\nonumber\frac{K^2_z}{{4a}A k_BT_e JN(0)}\int\frac{d{\bf q}'}{(2\pi)^2}\int\frac{d{\bf q}''}{(2\pi)^2}N_B(\beta_m\left(\epsilon({\bf q}')-\mu_m\right))\\
&\qquad\qquad\qquad\quad\times N_B(\beta_m\left(\epsilon({\bf q}'')-\mu_m\right))\frac{{\cal P}}{{\bf q}'\cdot{\bf q}''},\\
e_1&=\left[\frac{\zeta(3/2)}{(4\pi)^{3/2}}\right]\left(\frac{k_BT_e}{A}\right)^{1/2}\frac{K_zd_{\text{F}}}{A JN(0)a},\\
e_2&=\frac{{a^3}k_FK_zJS}{2\epsilon_0\epsilon_F}\int\frac{d{\bf q}'}{(2\pi)^2}{\cal P}\frac{\ln\left[\frac{1+\sqrt{1-\left(q'/2k_F\right)^2}}{q'/2k_F}\right]}{\left(1-\epsilon({\bf q}')/\epsilon_0\right)^2}.
\end{align}

These parameters show the effect from interfacial coupling, magnon-magnon interactions and thermal cloud on the magnon gap. Note that $e_2$ is proportional to the magnon-magnon interaction $K_z$ and the magnon-electron coupling $J$, thus representing the interference between magnons and electrons. In fact, the two-particle interaction occurs between magnons that are dressed by their coupling with electrons. Taking $\epsilon_0\sim K_z$ and the critical temperature of the ferromagnet $T_c \sim A/a^2k_B \ll T_e$ we estimate the dimensionless parameters as $e_0 \sim \left[1-\left(\frac{K_z}{A/a^2S}\right)\right]\left(\frac{K_z}{k_B T_e}\right)\left(\frac{\epsilon_F}{J}\right)$, $e_1 \sim  \left(\frac{T_e}{T_c}\right)^{1/2}\left(\frac{K_z}{k_BT_c}\right)\left(\frac{\epsilon_F}{J}\right)
\left(\frac{d_F}{a}\right)$ and  $e_2 \sim \left(\frac{J}{\epsilon_F}\right)$. For typical numbers we expect $e_2 \ll 1$, $e_0 =\mathcal{O}\left(1\right)$, but that $e_1$ can be rather large. 

\begin{figure}[h!]
\includegraphics[width=3.27in]{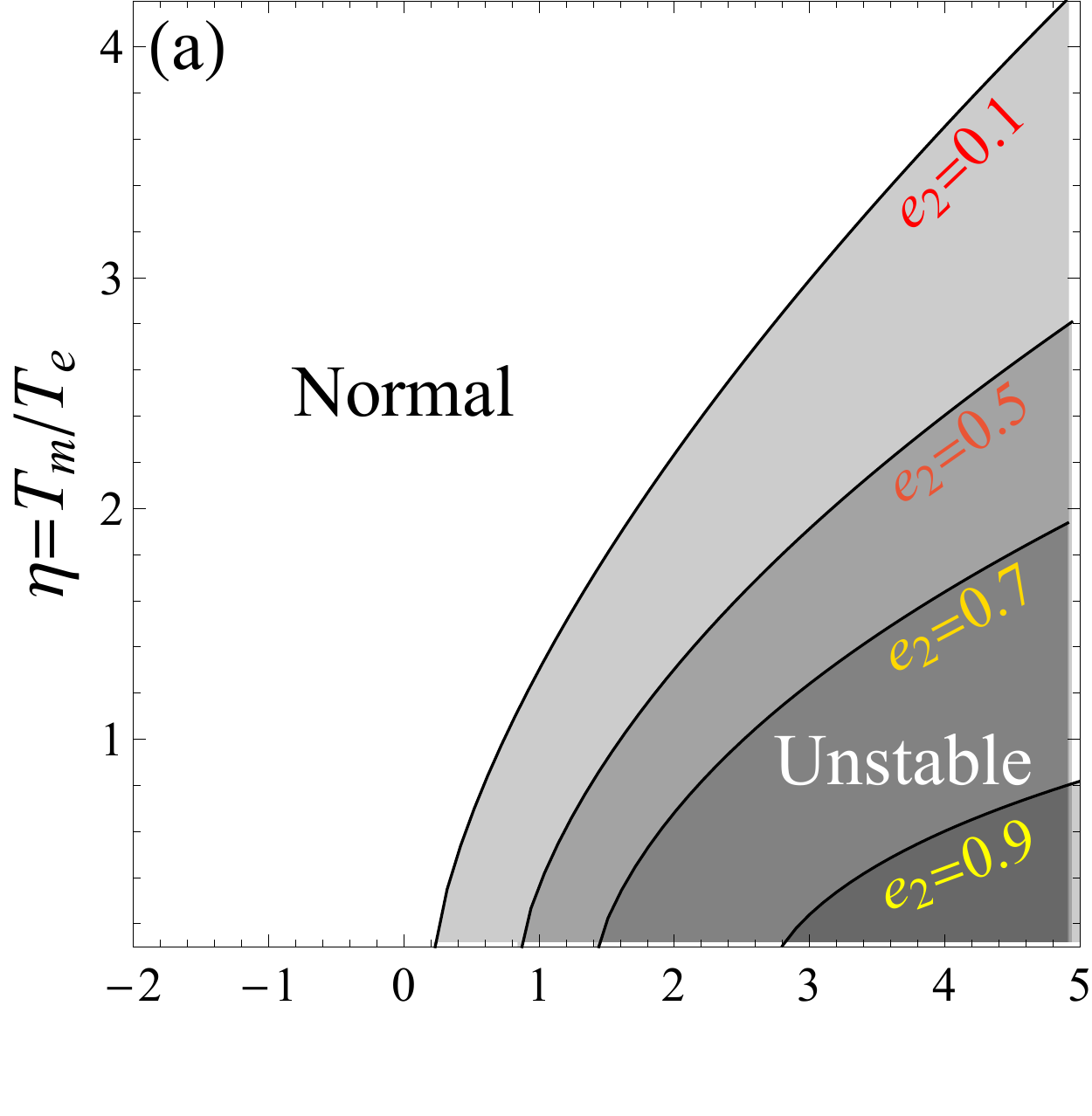}\vspace{-1.2cm}\\
\includegraphics[width=3.27in]{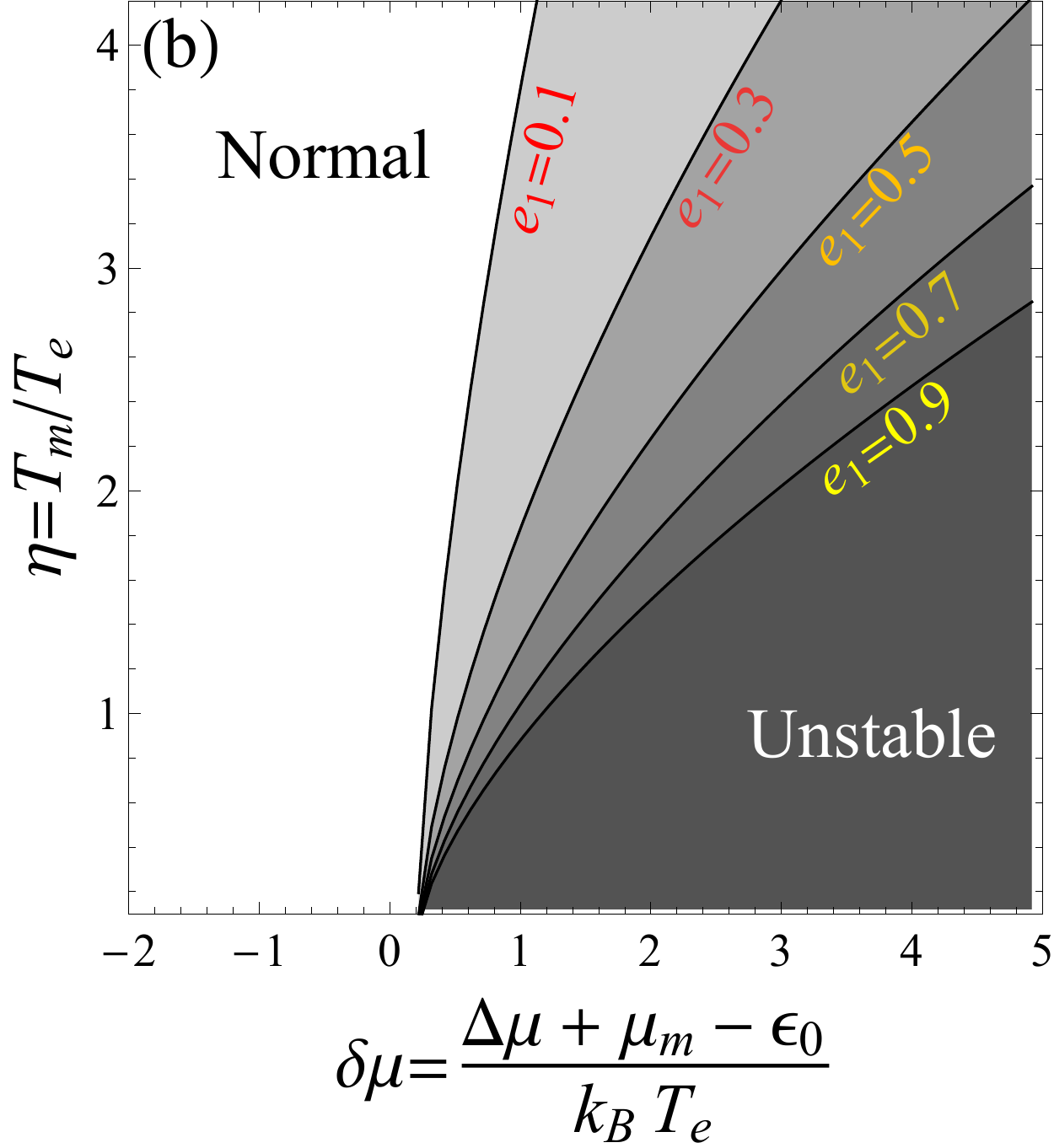}
\caption{Phase diagram for the stability of a driven magnon system as a function of the thermal imbalance $\eta$ and effective chemical potential $\delta\mu$ in units of the thermal energy $k_BT_e$. Each curve delimits both phases, being the right-hand side of the graph where the condition Eq. (\ref{eq:magnonBECcriterion}) is fulfilled and thus, corresponding to the unstable phase of magnons. In panel (a) we have $e_0=0.1$, $e_1=0.5$ and $e_2=0.1, 0.5, 0.7$ and  $0.9$. Meanwhile, in panel (b) $e_0=0.1$, $e_2=0.1$ and $e_1=0.1, 0.3, 0.5, 0.7$ and  $0.9$.}
		\label{fig:phasediag}
\end{figure}

The phase diagram is shown in Fig. \ref{fig:phasediag} as a function of the effective chemical potential $\delta\mu$ and thermal imbalance $\eta$, respectively. It consists of a stable and unstable magnon phase, separated by the line $F[\delta\mu,\eta;e_0,e_1,e_2]=0$, for different values of $e_2$ (panel (a)) and $e_1$ (panel (b)). It is worthwhile to note that in the limit $e_2\ll 1$ the criterion for instability reduces to that for a Bose-Einstein condensation in a Bose gas in the Popov approximation, for with the critical temperature is $\propto\left(\delta\mu-\epsilon_0\right)^{2/3}$. On the other hand, the unstable region diminishes as we increase both parameters $e_1$ and $e_2$. In particular, when $e_2>1$ the unstable region is suppressed. {This can be further analyzed by taking the limit $\eta\rightarrow 0$} in Eq. (\ref{eq:result_phasediagram}), where we see that $F\rightarrow e_0+(e_2-1)\delta\mu$, for large $\delta\mu$. Clearly, when $e_2>1$ the magnon gas never shows an instability if $e_0>0$. The term proportional to $e_2$ stems from combined effects of magnon-electron coupling and magnon-magnon interactions. We thus find that if these combined effects are sufficiently strong, an instability does not occur. While our perturbative approach is not valid in this regime, it still hints at interesting strong-coupling effects. Finally, we remark that at zero spin accumulation, $e_0$ and $e_2$ reflect the shift in the energy of single-particle ground state due to the electron-magnon coupling.

%%%%%%%%%%%%%%%%%%%%%%%Picture of the phase diagram
%%%%%%%%%%%%%%%%%%%%%%%%

%%%%%%%%%%%%%%%%%%%%%%%%%%%%%%%%%%%%
\section{Discussion and Conclusions}\label{sec5}
%%%%%%%%%%%%%%%%%%%%%%%%%%%%%%%%%%%%
In this article, we have presented a formalism to microscopically investigate how spin currents across an interface between a normal metal and a magnetic insulator lead to instabilities in the magnetic insulator. Our study has been based on a minimal microscopic model for the electrons, magnons in the magnetic insulator, and their coupling. 
While a direct connection between our results and experiments might be hard to establish, we do find that strong electron-magnon coupling and magnon-magnon interactions may prevent instabilities from occurring. This finding may be of relevance once interfacial electron-magnon interactions are being experimentally explored beyond the YIG-Pt paradigm.

Here, we have investigated electron-magnon and magnon-magnon interactions perturbatively. Future work should address the effects of strong interactions also by other means, i.e., by using the renormalization group. Another interesting direction is extending our theory beyond the linear stability analysis performed here to include the description of the dynamics in the unstable region of the phase diagram.

%%%%%%%%%%%%%%%%%%%%%%%
\begin{acknowledgments}
It is a pleasure to thank S. Bender, Y. Tserkovnyak and A. S. Nu\~nez for fruitful discussions. This work was supported by the Netherlands Organization for Scientific Research (NWO), and the European Research Council (ERC).
\end{acknowledgments}
%%%%%%%%%%%%%%%%%%%%%%%

\appendix

%%%%%%%%%%%%%%%%%%%%%%%%%%%%%%%%%%%%%%%%%%%%%%%%%%%%%%%%%%%
\section{Magnon self energy due to coupling with electrons}\label{sec:SelfEnergy}
%%%%%%%%%%%%%%%%%%%%%%%%%%%%%%%%%%%%%%%%%%%%%%%%%%%%%%%%%%%
In this appendix, we evaluate the self energy of the magnons due to their coupling with electrons. We start out with some general remarks for functions on the Keldysh contour. 

A function $F(t,t')$, whose arguments are defined on the Keldysh contour, can be decomposed into analytic parts by means of 
\begin{align}
F(t,t')=F^{\delta}(t)\delta(t,t')+\Theta\left(t,t'\right)F^{>}(t,t')+\Theta\left(t',t\right)F^{<}(t,t'),
\end{align}
with $\Theta\left(t,t'\right)$ the Heaviside step function on the Keldysh contour and $F^{\delta}(t)$ represents a possible $\delta$-singularity. The retarded and advanced components of $F(t,t')$ are related to the analytic parts by  
\begin{align}
F^{(\pm)}(t,t')=\pm\Theta\left(\pm(t,t')\right)\left(F^{>}(t,t')-F^{<}(t,t')\right),
\end{align}
where $\Theta\left(\pm(t,t')\right)\equiv\Theta\left(\pm(t-t')\right)$. We also have the Keldysh component
\begin{align}
F^{K}(t,t')=F^{>}(t,t')+F^{<}(t,t')
\end{align}
that typically is associated to the strength of fluctuations.

Applying the above definitions to Eq. (\ref{eq:selfenergygeneralized}), we see that the Fourier transform of the retarded (advanced) electron bubble is
\begin{align}\label{eq:electronbubblegeneral}
\Pi^{\sigma\sigma'}{}^{,(\pm)}({\bf k},\epsilon)=\nonumber &\int \frac{d\epsilon'}{(2\pi\hbar)}\int\frac{d\epsilon''}{(2\pi\hbar)}\sum_{{\bf k}''}A_{{\bf k}+{\bf k}''}(\epsilon')A_{{\bf k}''}(\epsilon'')\\
&\quad\times\frac{N_F\left(\epsilon'-\mu_{\sigma}\right)-N_F\left(\epsilon''-\mu_{\sigma'}\right)}{\epsilon'-\epsilon''-\epsilon^{\pm}}
\end{align}
where $A_{\bf k}(\epsilon)$ denotes the spectral function (being ${\bf k}$ a three-dimensional wavevector), $N_F (\epsilon)= [e^{\beta_e \epsilon}+1]^{-1}$ the Fermi distribution function, and $\mu_\sigma$ the chemical potential of electrons with spin projection $\sigma$. Ignoring electronic lifetime effects, we use  $A_{\bf k}(\epsilon)=2\pi\hbar\delta\left(\epsilon-\epsilon_{\bf k}\right)$ and approximate up to first order in spin accumulation. We then find that in the long-wavelength and small frequency limit the retarded (advanced) and Keldysh components of electron bubble at low temperatures read
\begin{align}\label{eq:selfenergyRAappendix}
\Pi^{\uparrow\downarrow}{}^{,(\pm)}({\bf k},\epsilon)\nonumber=&\frac{1}{2}N(0)a^3\left[\frac{1}{3}\left(\frac{{\bf k}}{2k_F}\right)^2-1\right]-\frac{N(0)a^3\epsilon\Delta\mu}{16\epsilon^2_F}\\
&\nonumber\times\left[\left(\frac{2k_F}{\bf k}\right)^2+1\right]\pm i \frac{\pi N(0)a^3k_F}{8\epsilon_F\left|{\bf k}\right|}\left(\Delta\mu-\epsilon\right)\\
&\qquad\qquad\qquad\qquad\times\Theta\left[1-\left(\frac{|{\bf k}|}{2k_F}\right)^2\right]
\end{align}
and
\begin{align}\label{eq:selfenergyKeldysh}
\Pi^{\uparrow\downarrow}{}^{,K}({\bf k},\epsilon)=-\frac{i\pi N(0)a^3 m k_B T_e}{\hbar^2k_F |{\bf k}|}\Theta\left[1-\left(\frac{|{\bf k}|}{2k_F}\right)^2\right],
\end{align}
with $a$ the lattice constant, $m$ the mass of electrons and $N(0)=mk_{F}/\pi^2\hbar^2$ the electronic density of states at the Fermi level. The imaginary part of the self-energy Eq. (\ref{eq:selfenergygeneralized}) represents the rate of change of the number of magnons. From Eq. (\ref{eq:selfenergyRA}), we see that the evolution of the number of magnons corresponds to the competition between the spin transfer ($\propto \Delta \mu \equiv \mu_\uparrow-\mu_\downarrow$) and spin pumping mechanisms ($\propto \epsilon$). In principle, the magnon-electron coupling matrix element can be determined in terms of the mixing conductance, but this identification will not be pursued here \cite{Tserkovnyak}. 

\section{Ladder approximation}\label{sec:LadderApproximation}
%%%%
In section III A, we derived an action for the gas of magnons, which are excited by the combined effects of a thermal gradient and the spin-transfer torque across the F$|$M interface. Here, we discuss more details of this derivation. 

We introduce the order parameter $\langle\phi_{\bf q}(t)\rangle$, that characterizes the instability. This is accomplished by performing a Legendre transformation \cite{Amit} on the magnon field variables that ultimately leads to an effective action for $\psi_{\bf q}(t)\equiv\langle\phi_{\bf q}(t)\rangle$. With this aim we start out by introducing the generating functional for Keldysh Green's functions, following Ref.~\cite{stoof1}, as
\begin{align}\label{partfunc}
{\cal Z}[J,{J}^*]=\int {\cal D}[\phi^*]{\cal D}[\phi]\exp\left\{\frac{i}{\hbar}{\cal S}_{m}[\phi,\phi^*]+i\left(J^{\alpha}\phi_{\alpha}+c.c.\right)\right\}
\end{align}
where $J_{\alpha}$ and $J^{\alpha}$ are sources that are defined on the Keldysh contour and summation over repeated indices means an integration over space and time coordinates. Then the Legendre transformation, 
\begin{align}
\Gamma[\psi,\psi^*]\equiv\psi^{\alpha}J_{\alpha}+J^{\alpha}\psi_{\alpha} -W[J,J^*]
\end{align}
where $W[J,J^*]=-i \log {\cal Z}[J,J^*]$ is the generating functional for connected Green's functions, which can be evaluated in terms of a perturbation series. The order parameter is then $ \psi_{\alpha}\equiv \delta W/\delta J^{\alpha}=\langle\phi_{\alpha}\rangle$. The functional $-\hbar\Gamma[\phi,\phi^*]$, that generates all one-particle irreducible diagrams, corresponds to the effective action. We now do perturbation theory in the interaction to evaluate the effective action 
in terms of the coefficients $\Gamma^{(2n)}$ (up to $n=2$) of the term that is of order $n$ in the fields. Those coefficients in momentum space correspond to
\begin{align}\label{eq:gamma2}
\hbar\Gamma^{(2)}({\bf q};t,t')=\nonumber &-\left[\left(i\hbar\frac{\partial}{\partial t}-\epsilon\left({\bf q}\right)+\mu_m\right)\delta(t,t')\right.\\
&\left.\qquad-\hbar\Sigma_{\text{st}}({\bf q};t,t')\right]+\hbar\Sigma_{\text{in}}({\bf q};t,t'),
\end{align}
\begin{align}\label{eq:gamma4}
\hbar\Gamma^{(4)}\left({\bf q},{\bf q}',{\bf Q};t,t'\right)=&\text{T}\left({\bf q},{\bf q}',{\bf Q};t,t'\right)+\text{T}\left(-{\bf q},{\bf q}',{\bf Q};t,t'\right)
\end{align}
where the interacting self-energy $\hbar\Sigma_{\text{in}}$ was introduced by the Dyson equation for the magnon Green's function ${\cal G}={\cal G}_0+{\cal G}_0\hbar\Sigma_{\text{in}}{\cal G}$, with ${\cal G}_0$ the dressed magnon propagator due to the contact with the normal metal, see Fig. \ref{fig:DiagrammaticRepresentation}(a), according to Eq. (\ref{eq:selfenergygeneralized}). The two-body interaction is evaluated within the T-matrix (or ladder) approximation \cite{stoof1}, diagrammatically indicated in Fig. \ref{fig:DiagrammaticRepresentation}(c), which describes the scattering of two magnons that at time $t'$ have the momenta $\hbar({\bf Q}\pm{\bf q}')$ and at time $t$ the momenta $\hbar({\bf Q}\pm{\bf q})$. The exact interacting self-energy obeys the relation 
\begin{align}\label{eq:interactingselfenergy}
\hbar\Sigma_{\text{in}}({\bf q};t,t')=i \int\frac{d{\bf q}'}{(2\pi)^2}\hbar\Gamma^{(4)}\nonumber({\bf q}-{\bf q}&',{\bf q}-{\bf q}',{\bf q}+{\bf q}';t,t')\\
&\times{\cal G}({\bf q}';t',t).
\end{align}
In Fourier space the expression for the retarded component of the interacting self-energy Eq. (\ref{eq:interactingselfenergy}) take the form
\begin{widetext}
	\begin{align}\label{eq:InteractingSelfEnergy1}
	\hbar\Sigma^{(+)}_{\text{in}}({\bf q};\epsilon)=
	i\int\frac{d{\bf q}'}{(2\pi)^2}\int\frac{d\epsilon'}{(2\pi)}\nonumber\Gamma^{(+)}_{4}&\left({\bf q}-{\bf q}',{\bf q}-{\bf q}',{\bf q}+{\bf q}';\epsilon+\epsilon'
	\right){\cal G}^{(+)}({\bf q}';\epsilon'){\hbar\Sigma}^{<}_{\text{st}}({\bf q}';\epsilon'){\cal G}^{(-)}({\bf q}';\epsilon')\\
	&+i\int\frac{d{\bf q}'}{(2\pi)^2}\int\frac{d\epsilon'}{(2\pi)}\Gamma^{<}_{4}\left({\bf q}-{\bf q}',{\bf q}-{\bf q}',{\bf q}+{\bf q}';\epsilon+\epsilon'
	\right){\cal G}^{(-)}({\bf q}';\epsilon').
	\end{align}
\end{widetext}
The evaluation of Eq. (\ref{eq:InteractingSelfEnergy1}) is carried out by expanding up to second order in the coupling with the leads. In this approach the various components of the Green's function for magnons are approximated by 
\begin{align}\label{eq: greenexpansion}
{\cal G}^{(\pm)}=&{\cal G}_{0}^{(\pm)}+{\cal G}_{0}^{(\pm)}\hbar\Sigma^{(\pm)}_{\text{st}}{\cal G}_{0}^{(\pm)}+\dots\\
{\cal G}^{<}=&{\cal G}^{(+)}\hbar\Sigma^{<}_{\text{st}}{\cal G}^{(-)}={\cal G}^{(+)}_0\hbar\Sigma^{<}_{\text{st}}{\cal G}_0^{(-)}+\dots
\end{align}
as can be seen in Fig. \ref{fig:DiagrammaticRepresentation}(b). On the other hand, the magnon-magnon interactions will be approximated by a contact interaction, and therefore  $\text{T}^{(\pm)}({\bf q},{\bf q}',{\bf Q};\epsilon)\approx \frac{K_z}{2}$. After some manipulations \cite{stoof1}, we arrive at the semiclassical effective action Eq. (\ref{eq:effAction}), describing the low-energy dynamics of the interacting magnons.

 %%%%%%
 \section{Self-consistent relations: chemical potentials and temperatures}\label{sec:SelfConsistent}
 %%%%%%
 
 In this section we compute the chemical potential and temperature of magnons assuming that magnons are sufficiently close to equilibrium. 

 The total spin-current flowing across the interface is quantified by the rate of change of magnons in the ferromagnet, that may be obtained following standard methods described in Ref. [\onlinecite{stoof1}]. It consists of analyzing the stochastic dynamics of magnons due to the coupling with the metal, that ultimately turns out in a Boltzmann equation. For this purpose we split the magnon field, in Eq. (\ref{eq:effectiveaction}), into semiclassical and fluctuating parts according to $\phi_{\bf q}(t_{\pm})=\varphi_{\bf q}(t)\pm\xi_{\bf q}(t)/2$, where $t_{\pm}$ refers to the forward and backward branches of the Keldysh contour, respectively and $\xi_{\bf q}(t)$ the fluctuations. After integrating out the fluctuations $\xi_{\bf q}(t)$ in the action, Eq. (\ref{eq:effectiveaction}), we find that the field $\varphi_{\bf q}(t)$ obeys the Langevin equations
 \begin{align}\label{eq:langevinmagnoneq}
 i\hbar\frac{\partial}{\partial t}\varphi_{\bf q}(t)=(\epsilon &({\bf q})\nonumber-\mu_m)\varphi_{\bf q}(t)\\
 &+\int dt' \hbar\Sigma^{(+)}_{\text{st}}({\bf q};t,t')\varphi_{{\bf q}}(t')+\eta_{{\bf q}}(t)
 \end{align}
 and
 \begin{align}\label{eq:langevinmagnoneqc}
 -i\hbar\frac{\partial}{\partial t}\varphi^{*}_{\bf q}(t)=&(\epsilon \nonumber({\bf q})-\mu_m)\varphi^{*}_{\bf q}(t)\\
 &+\int dt' \varphi^{*}_{{\bf q}}(t')\hbar\Sigma^{(-)}_{\text{st}}({\bf q};t,t')+\eta^{*}_{{\bf q}}(t)
 \end{align}
 with the Gaussian stochastic noise $\eta_{{\bf q}}(t)$ and $\eta^{*}_{{\bf q}}(t)$ is zero on average and has the correlations
 \begin{align}
 \langle\eta_{{\bf q}}(t)\eta^{*}_{{\bf q} }(t' )\rangle=\frac{i\hbar}{2}\hbar\Sigma^{K}_{\text{st}}({\bf q};t-t')~.
 \end{align}
 In the low-energy approximation we see that the strength of the noise is evaluated directly from the combination of Eq. (\ref{eq:selfenergygeneralized}) and Eq. (\ref{eq:selfenergyKeldysh}). This relation between noise and damping stems from the fluctuation-dissipation theorem \cite{L.P. Kadanoff} and ensures us that the magnon gas relaxes to thermal equilibrium.
 
Note that to obtain Eq. (\ref{eq:langevinmagnoneq}) and (\ref{eq:langevinmagnoneqc}), as a first step, we have not taken into account the interaction between magnons. However, the collision terms will be included next in the Boltzmann equation. We take into account the leading low-energy contribution of the self-energy, i.e. $\int dt' \hbar\Sigma^{(+)}_{\text{st}}({\bf q};t,t')\varphi_{{\bf q}}(t')\simeq \left[\hbar\Sigma^{(+)}_{\text{st}}({\bf q},0)+\hbar\Sigma^{(+)}{}'_{\text{st}}({\bf q},0)i\hbar\frac{\partial}{\partial t}\right]\varphi_{{\bf q}}(t)$. Finally, the rate equation for the magnons due to the coupling at the interface with the electron reservoir is written explicitly as
\begin{widetext}
\begin{align}\label{eq:magnonrateST}
i\hbar&\left.\left(\frac{\partial n({\bf q},t) }{\partial t}\right)\right|_{\text{st}}=2i\text{Im}\left[\hbar\Sigma^{(+)}{}'_{\text{st}}({\bf q},0)\right]\left(\epsilon_{\bf q}-\mu_m\right)n({\bf q},t)+2i\text{Im}\left[\hbar\Sigma^{(+)}_{\text{st}}({\bf q},0)\right]n({\bf q},t)-\frac{1}{2}\hbar\Sigma^{K}_{\text{st}}({\bf q},0)
\end{align}
\end{widetext}
with $n({\bf q},t)=\langle \phi^{*}_{{\bf q}}\phi_{{\bf q}}\rangle$ and $\langle\dots\rangle$ stand for averaging over noise realization. Therefore, the full dynamics for the distribution of magnons in the ferromagnet is determined by the Boltzmann equation 
\begin{align}\label{eq:boltzman}
\frac{\partial n({\bf q},t)}{\partial t}=-2\alpha\omega_{\bf q}&n({\bf q},t)+\left.\left(\frac{\partial n({\bf q},t) }{\partial t}\right)\right|_{\text{st}}
\end{align}
with $\alpha$ the Gilbert damping contant and where the collisions terms has been considered.  Taking moments of Eq. (\ref{eq:boltzman}) we obtain a closed set of equations for the total number of magnons and energy. In equilibrium, there will be neither spin flow nor energy transfer through the interface. This is implemented by requiring $\partial n(t)/\partial t\equiv\frac{\partial}{\partial t}\sum_{\bf q} n({\bf q},t)=0$ and $\partial \epsilon(t)/\partial t\equiv\frac{\partial}{\partial t}\sum_{\bf q}\epsilon_{\bf q}n({\bf q},t)=0$, that turn out in the pair of Eqs. (\ref{eq:criticalSAandT(a)}) and (\ref{eq:criticalSAandT(b)}).

\bibliographystyle{elsarticle-num}

\end{document}